\documentclass[twocolumn,showpacs,preprintnumbers]{revtex4}%
\usepackage{amsfonts}
\usepackage{amsmath,epsfig}
\usepackage{graphicx}
\usepackage{dcolumn}
\usepackage{bm}
\usepackage{amssymb}
\usepackage{amsmath}%

\begin{document}
\title{Combustion of a neutron star into a strange quark star: The neutrino signal}
\author{Giuseppe Pagliara$^{\text{(a)}}$, Matthias
  Herzog$^{\text{(b)}}$ and Friedrich K.~R\"opke$^{\text{(c)}}$}
\affiliation{$^{\text{(a)}}$Dipartimento di Fisica e Scienze della Terra dell'Universit\`a di Ferrara and INFN
Sezione di Ferrara, Via Saragat 1, I-44100 Ferrara, Italy}
\affiliation{$^{\text{(b)}}$Max-Planck-Institut f{\"u}r Astrophysik,
  Karl-Schwarzschild-Strasse~1, D-85741 Garching, Germany}
\affiliation{$^{\text{(c)}}$Institut f{\"u}r Theoretische Physik und Astrophysik, Universit\"at W\"urzburg,
  Emil Fischer-Strasse~31, D-97074 W{\"u}rzburg, Germany}

\begin{abstract}
There are strong indications that the process of conversion of a
neutron star into a strange quark star proceeds as a strong
deflagration implying that in a few milliseconds almost the whole star
is converted. Starting from the three-dimensional hydrodynamic
simulations of the combustion process which provide the temperature
profiles inside the newly born strange star, we calculate for the
first time the neutrino signal that is to be expected if such a
conversion process takes place. The neutrino emission is characterized
by a luminosity and a duration that is typical for the signal expected
from protoneutron stars and represents therefore a powerful source of
neutrinos which could be possibly directly detected in case of events
occurring close to our Galaxy.  We discuss moreover possible
connections between the birth of strange stars and explosive phenomena
such as supernovae and gamma-ray-bursts.
\end{abstract}

\maketitle

\section{Introduction}
Under the hypothesis that the true ground state of nuclear matter is
cold strange quark matter
\cite{Itoh:1970uw,Bodmer:1971we,1984PhRvD..30..272W,Haensel:1986qb,Alcock:1986hz} it is likely that
compact stars exist that are composed almost entirely of quark matter, the 
so-called strange stars. In turn, the birth of these stellar objects can
be thought of as a decay of a metastable neutron star into a more
bound configuration with a consequent huge release of energy, of the
order of $10^{53}$ erg. If released in a short time period, the
fascinating possibility exists that this energy source is associated
with the most extreme explosions of the Universe, i.e. supernovae (SNe)
and gamma-ray-bursts (GRBs). The strange quark matter hypothesis could possibly 
be proven also by detecting the gravitational wave signal associated 
with the merger process of two strange stars as pointed out in recent 
simulations \cite{Bauswein:2008gx,Bauswein:2009im}.

There are several issues related to the conversion process that have
been investigated in the past.  A first problem concerns the initial
seed of strange quark matter which then triggers the conversion:
Several studies have addressed the thermal and quantum nucleation of
the first drop of quark matter, and different possible scenarios have
been proposed
\cite{Olesen:1993ek,Iida:1998pi,Berezhiani:2002ks,Drago:2004vu,Bombaci:2004mt,Bombaci:2006cs,Mintz:2009ay,Bombaci:2009jt,Logoteta:2012ms}.
A second and more complicated question concerns the time needed for
the conversion of the whole star. In the first study addressing this
issue \cite{1987PhLB..192...71O}, a laminar conversion front was
assumed with a resulting time needed for the conversion of up to $100$s.
Such a slow process would provide a very faint neutrino signal and
thus not be relevant from the phenomenological point of view.  On the
other hand, it was pointed out and estimated that hydrodynamical
instabilities can substantially increase the velocity of the
conversion front \cite{Horvath:1988nb,Lugones:2002vj} leading to a regime, in most of
the cases, of strong deflagration \cite{Cho:1993de,Drago:2005yj}. In
Ref. \cite{Drago:2005yj}, in particular, many different equations of
state (including hyperons, color superconductivity and mixed phases)
have been tested, also considering the effects of keeping the flavor 
composition of matter fixed during the combustion, and no case has
been found in which the conversion is so rapid to turn itself into a
detonation process. Also, the temperature inside the newly born star
has been estimated with resulting central temperatures of the order of
tens of MeV. 
Finally, a numerical study on this problem has recently
provided more quantitative results on the conversion
process \cite{Herzog:2011sn}: Three-dimensional
hydrodynamic simulations have been conducted under the hypothesis 
that the conversion of a hadronic
neutron star into a strange quark star is a combustion process. The
combustion turned out to be turbulent due to the growth of buoyancy
instabilities. Turbulence enhances the burning velocity
considerably, leading to short conversion time scales of a few
milliseconds. Moreover, it has been found that, even if strange quark matter is
absolutely stable, the hydrodynamic combustion process is not able to burn the
whole neutron star; a sizable layer survives which then probably
converts on a longer time scale.

Finally, a third problem that has never been addressed in detail
concerns how the huge energy of the conversion process is emitted by
the star. Clearly, the heat released by the formation of strange quark
matter is dissipated via a strong neutrino emission, very similar to
the emission of a protoneutron star. In this paper, starting from the
hydrodynamic simulations of the conversion process, we compute for the
first time the neutrino signal that is to be expected if a neutron
star decays into a strange quark star (some simple estimates have been
presented in \cite{Keranen:2004vj}). In particula,r we present results
for the diffusion of the neutrinos inside the newly born quark star
and the consequent cooling of the star. As expected, both the
luminosity and the duration of the neutrino signal are very similar to
the ones of protoneutron stars. The birth of a strange star is thus
phenomenologically very relevant, because the corresponding neutrino
signal could be detected by the presently available neutrino
telescopes if it occurs close enough to our Galaxy. Moreover, the
formation of strange quark matter in a magnetar could represent a
possible extension of the so-called protomagnetar model of GRBs
\cite{Metzger:2010pp} which, as we will discuss, might be able to
explain some recent puzzling observations of GRBs.

The paper is organized as follows: In Sec. II we discuss the equations
of state (EOSs) adopted for nucleonic matter and for strange quark
matter. In particular, we will consider EOSs for strange quark matter
that allow us to obtain a maximum mass for cold and catalyzed stellar
configurations compatible with the $2M_{\odot}$ limit
\cite{Demorest:2010bx}.  In Sec. III, we review the procedure adopted to
simulate the conversion process as done in Ref.~\cite{Herzog:2011sn},
and we show the initial conditions for the neutrino diffusion calculations. In
Sec. IV, we present and discuss the calculation of the process of
diffusion of neutrinos within the newly born strange quark star. In
Sec. V, after a discussion on the possible phenomenological implications
of the process under study, we draw our conclusions.

\section{Equations of state}
For the nucleonic matter EOS we use the table computed by Lattimer and
Swesty (LS EOS) \cite{lattimer1991a}, which is based on a liquid drop
model. To be consistent with the observation of a two solar mass (see
below) neutron star, we apply an incompressibility modulus of $K=220$
MeV. For reasons discussed in detail in \cite{Herzog:2011sn},
nucleonic matter EOSs of much higher stiffness, e.g. the relativistic
mean field EOS by \cite{shen1998a}, could not be employed, because they do not 
allow for exothermic combustion.
For the strange quark matter EOS we use, as customary, the MIT bag model with
the inclusion of the perturbative QCD corrections \cite{Farhi:1984qu,Fraga:2001id}. For this work, we
neglect the effects related to the appearance of diquark condensates,
as in the color-flavor-locking phase, for instance \cite{Alford:2007xm}. A detailed study
on this possibility will be important: A second order phase transition
from unpaired quark matter to gapped matter could indeed take place in
the astrophysical object that we are considering, possibly providing
interesting signatures on the neutrino signal \cite{Carter:2000xf}.
Also, the change of the chemical composition of the star during deleptonization 
could lead to first-order phase transition between different quark phases
\cite{Pagliara:2010na}.
Concerning the free parameters of the quark matter model, the
discovery of a $2 M_{\odot}$ compact star \cite{Demorest:2010bx} has
significantly reduced its parameter space as shown in
\cite{Weissenborn:2011qu}. We consider two sets of parameters, taken
from \cite{Weissenborn:2011qu}, both allowing us to reach a maximum mass
for strange quark stars of $2 M_{\odot}$. We fix the current mass of the
strange quark to $m_s=100$ MeV, and we then consider $B_{\mathrm{eff}}^{1/4}=142$MeV -
$a_4=0.9$ (set1) and $B_{\mathrm{eff}}^{1/4}=141$MeV - $a_4=0.65$ (set2) where
$B_{\mathrm{eff}}$ and $a_4$ represent the effective bag constant and the
coefficient of the $\mu^4$ term in the pressure of the quark phase, 
respectively ($\mu$ is the quark chemical potential), as in
Ref. \cite{Alford:2004pf}.  While the two sets provide the same
maximum mass, they correspond to different values of the energy per
baryon of the ground state of strange quark matter: for set1 $E/A= 860$ MeV
and for set2 $E/A=930$ MeV; in turn, this implies that for set1 the EOS
is soft and for set2 stiff. Indeed, these two sets of parameters lie on
the two-flavor and three-flavor lines, respectively, in Fig. 1 of
\cite{Weissenborn:2011qu}. 
The EOSs are expressed as tables with the 
baryon density and the temperature as independent variables.
Finally, the chemical potentials of quarks and electrons
are fixed by the beta equilibrium conditions. Since 
our starting configuration is a cold and catalyzed neutron star,
we fix to zero the chemical potential of neutrinos (see discussion in the following).

\section{Combustion simulations and initial conditions for the cooling}
To simulate the combustion process we adopt the scheme already used in
\cite{Herzog:2011sn}.  We solve the Euler equations by using a well-tested grid code that
employs a finite volume discretization, the so-called 
piecewise parabolic method \cite{Colella:1982ee}. In this code, the conversion is described in a
discontinuity approximation meaning that the hadronic and newly
converted strange
quark matter is separated by a {\it conversion front}.
This conversion front is
modeled by using a level-set method \cite{reinecke1999a,reinecke2002d,roepke2005b}. Further references are
given in \cite{Herzog:2011sn}. In our simulations, general relativity 
effects are included only by using an effective relativistic gravitational potential
based on the Tolman-Oppenheimer-Volkov equations. It has been estimated
in Ref.\cite{Bhattacharyya:2007dt} that general relativity effects 
and the rotation of the star could lead to qualitative differences in the combustion.  
Presently, a 3D hydrodynamic simulation including also the before-mentioned effects
is computationally quite challenging, but it is of course 
an important outlook for our study.

\begin{table}
\begin{center}
    \begin{tabular}{ | l | l | l | l | l | l |}
    \hline
    $M_\odot$ (set 1) & a & b & c & d & e \\ \hline
    1.4 & 10.7 & 0.048 & 0.0025 & 39.1 & 18.4 \\ \hline
    1.8 & 9.5 & 0.054 & 0.0023 & 43.8 & 18.9 \\ \hline
    \end{tabular}
\caption{Parameters of the fit to the numerical results of temperature profiles.}
\end{center}
\end{table}

The conversion process is triggered by a strange quark matter seed in
the center of a hadronic neutron star. Numerically, at the start of
the simulation a volume in the center is converted instantly; the
conversion front subsequently propagates outwards as a deflagration
wave.  The initial, laminar conversion velocity is based on detailed
flame calculations of \cite{niebergal2010a}. Turbulent conversion
velocities are calculated by means of a sophisticated subgrid scale
turbulence model \cite{schmidt2006b,schmidt2006c}.

Except for the EOS for strange quark matter, we use exactly the same
initial setup as in \cite{Herzog:2011sn}.
We simulate one octant of a neutron star in three dimensions and apply
a resolution of 128 grid cells per dimension on a moving hybrid grid
which provides optimal resolution in the regions of interest. This
resolution is sufficient for our purpose, as was discussed in
\cite{Herzog:2011sn}, where also further details of our initial setup can be found.

\begin{figure*}
  \centering
  \begin{minipage}[b]{0.45\linewidth}\centering
    \includegraphics[width=\linewidth]{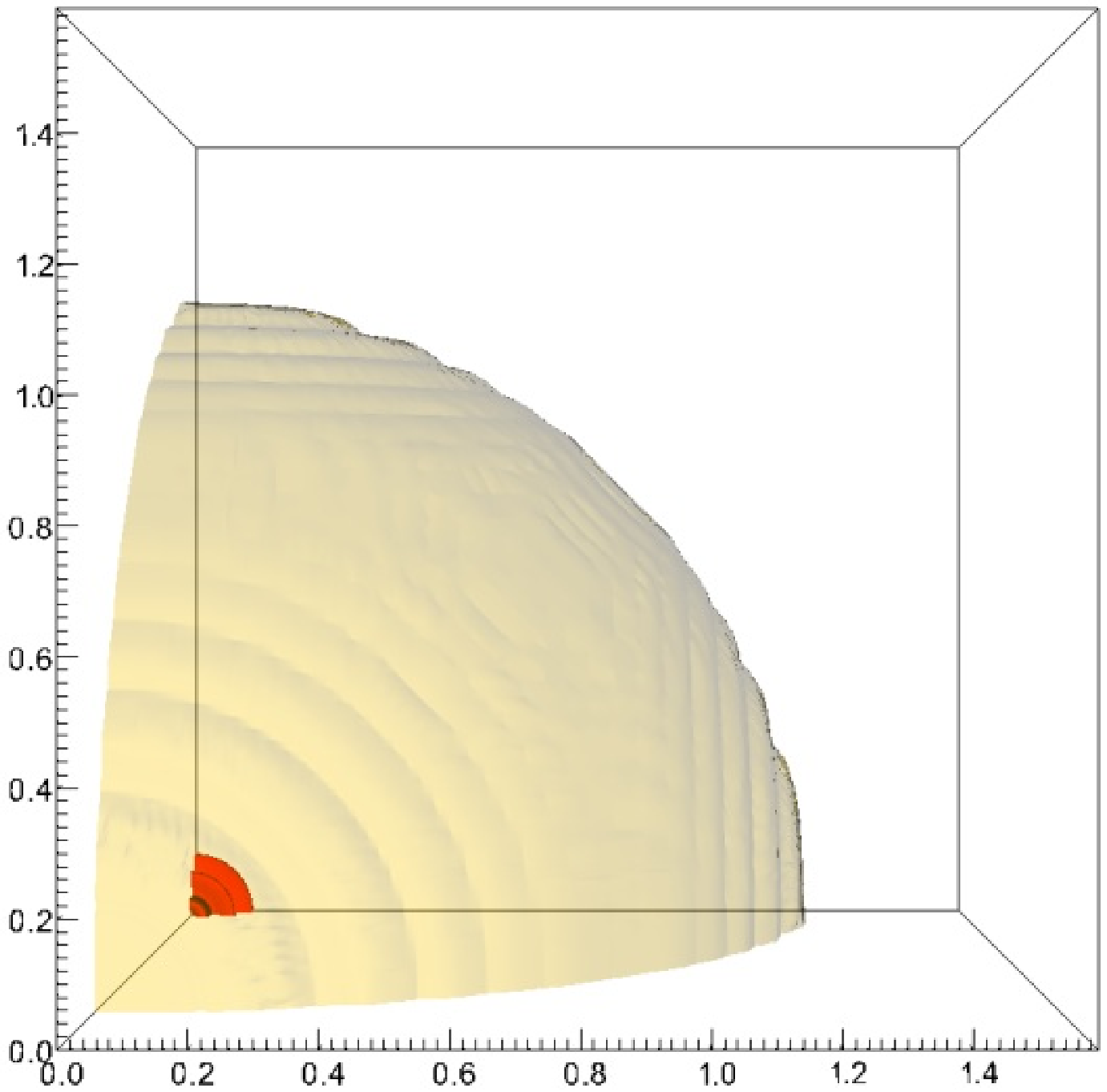}\\
    (a) $t = 0$
 \vspace{1.6cm}
  \end{minipage}
  \hspace{0.08\linewidth}
  \begin{minipage}[b]{0.45\linewidth}\centering
    \includegraphics[width=\linewidth]{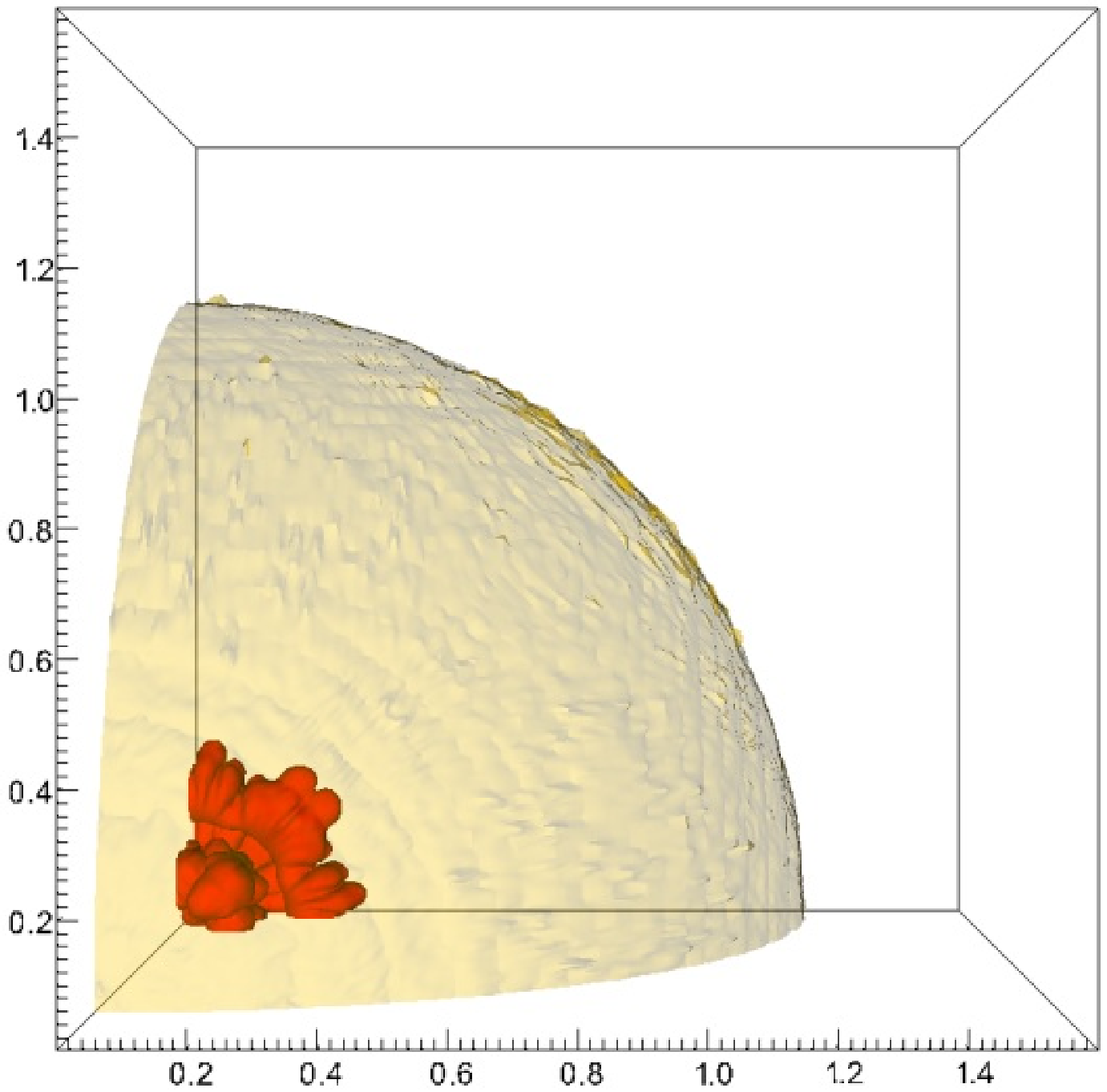}\\
    (b) $t = 0.7\, \mathrm{ms}$
    \vspace{1.6cm}
  \end{minipage}

  \begin{minipage}[b]{0.45\linewidth}\centering
    \includegraphics[width=\linewidth]{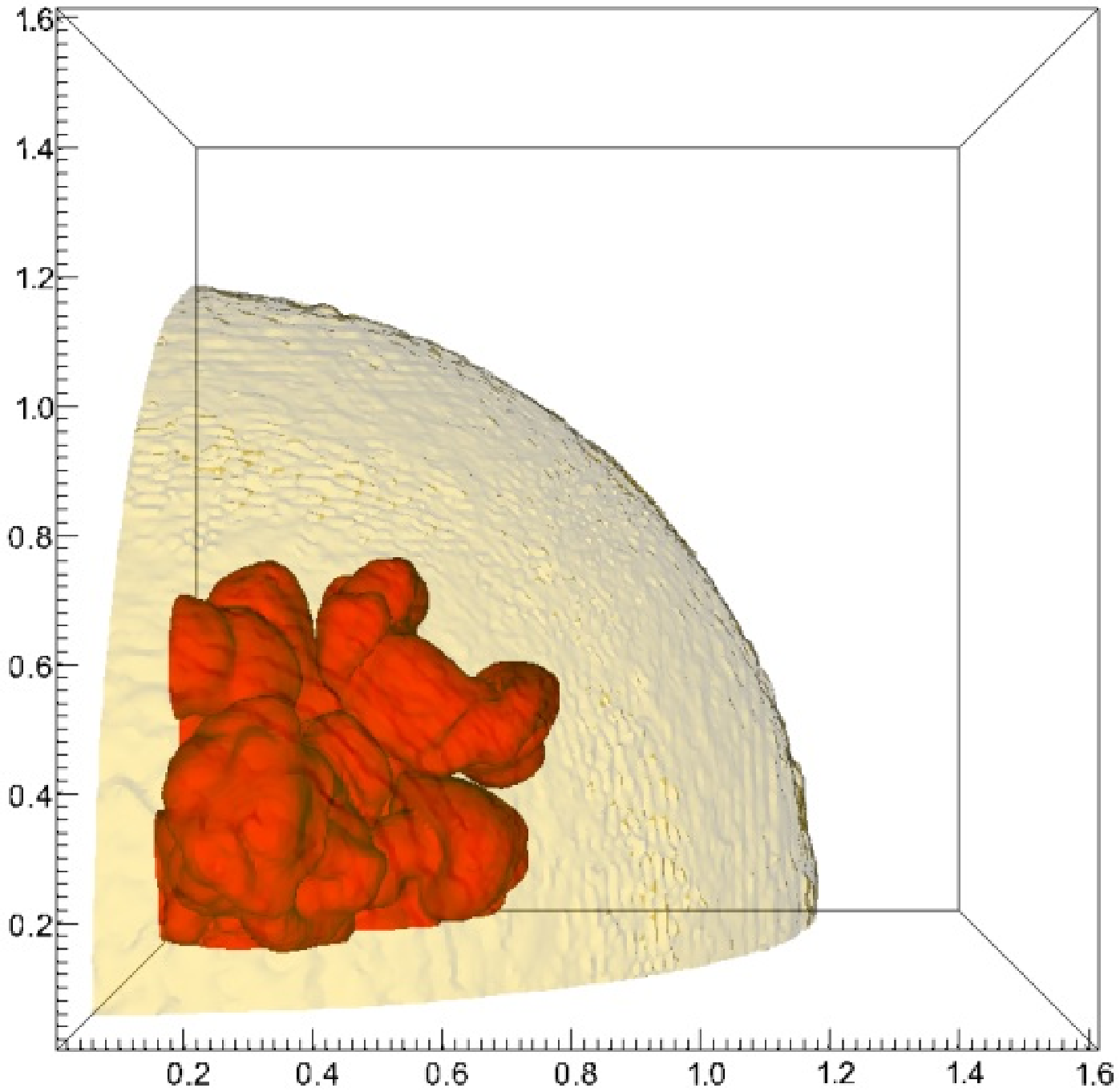}\\
    (c) $t = 1.2\, \mathrm{ms}$
  \end{minipage}
  \hspace{0.08\linewidth}
  \begin{minipage}[b]{0.45\linewidth}\centering
    \includegraphics[width=\linewidth]{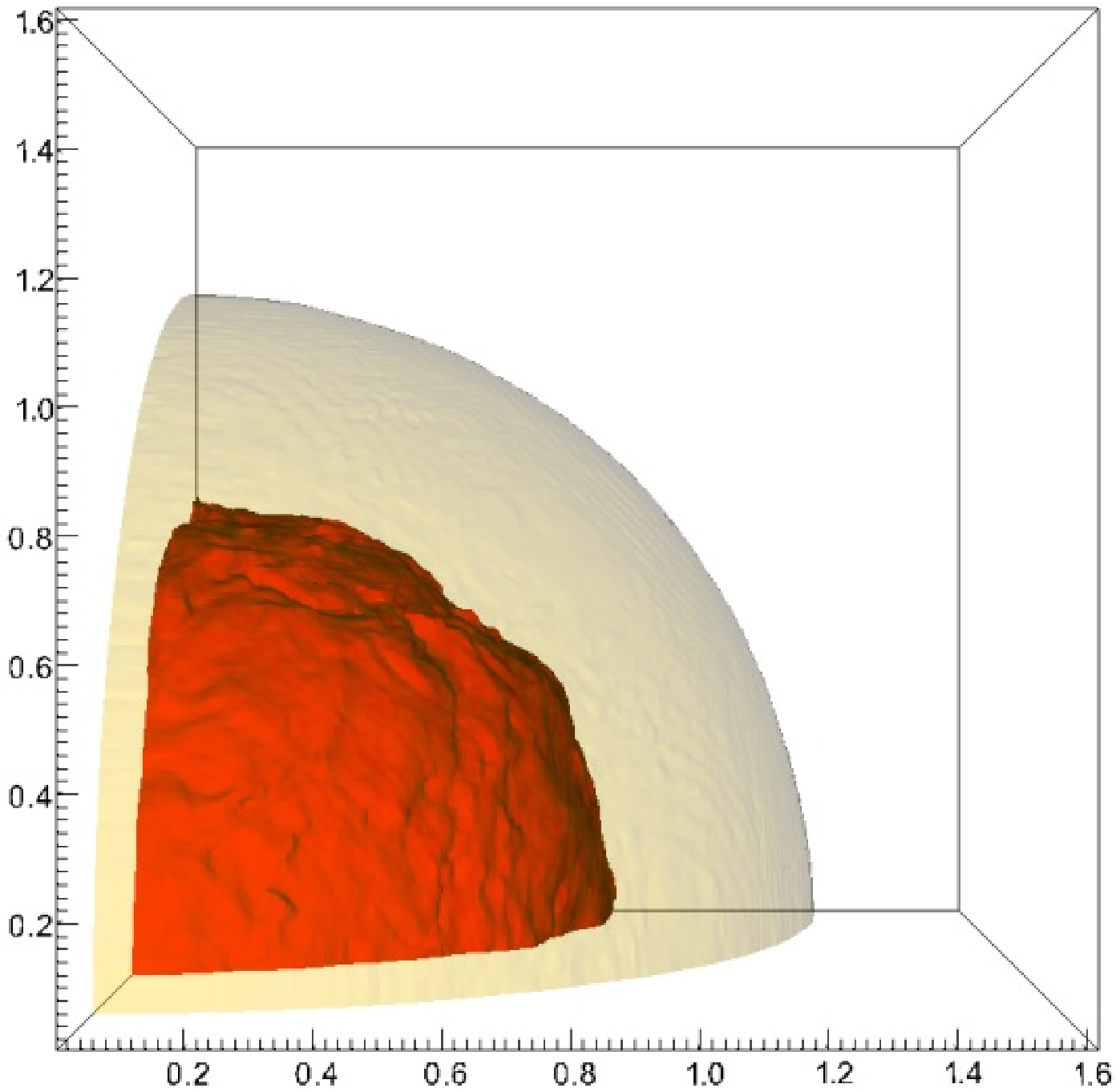}\\
    (d) $t = 4.0\, \mathrm{ms}$
  \end{minipage}

  \caption{(color online) Model: Set 1, $M = 1.4 M_{\odot}$. 
Conversion front (red) and
    surface of the neutron star (yellow) at different times $t$. Spatial units
    $10^6\,\mathrm{cm}$. }
\end{figure*}

We have performed the simulations of the combustion by using the
previously described strange quark matter and nucleonic matter EOSs and for two values of the mass
of the initial neutron star configuration: $1.4$ and $1.8 M_{\odot}$.

In Fig.~1, we illustrate the conversion process with snapshots taken at
four different instants in time for a $1.4 M_{\odot}$ neutron star
(set1).  While the soft EOS obtained with set1 allows for a
quite successful conversion of the star, for the stiff EOS
  obtained with set2, the requirement of having an exothermic process
does not allow the combustion. The neutron star cannot be converted --
at least within our framework.

In Fig. 2, we show the
results for the temperature profiles as a function of pressure 
obtained from the combustion simulations. At the center of the star the temperature 
reaches quite high values, $40-50$ MeV, and it drops steeply at the interface between the
burned material and the unburned one.
Together with the numerical
results, indicated by dots and boxes, in Fig. 2 we show parametrizations of the
temperature profiles that we will use as initial conditions for the
neutrino diffusion calculation. The parametrization reads 
$T=a \mathrm{Arctan}((P - b)/c)+dP+e$, where $T$ and $P$ are 
temperature and pressure, respectively, and the parameters
$a,b,c,d,e$ are obtained by fitting the numerical results (see
Table 1 for their numerical values).

\begin{figure}[ptb]
\vskip 0.5cm
\begin{centering}
\epsfig{file=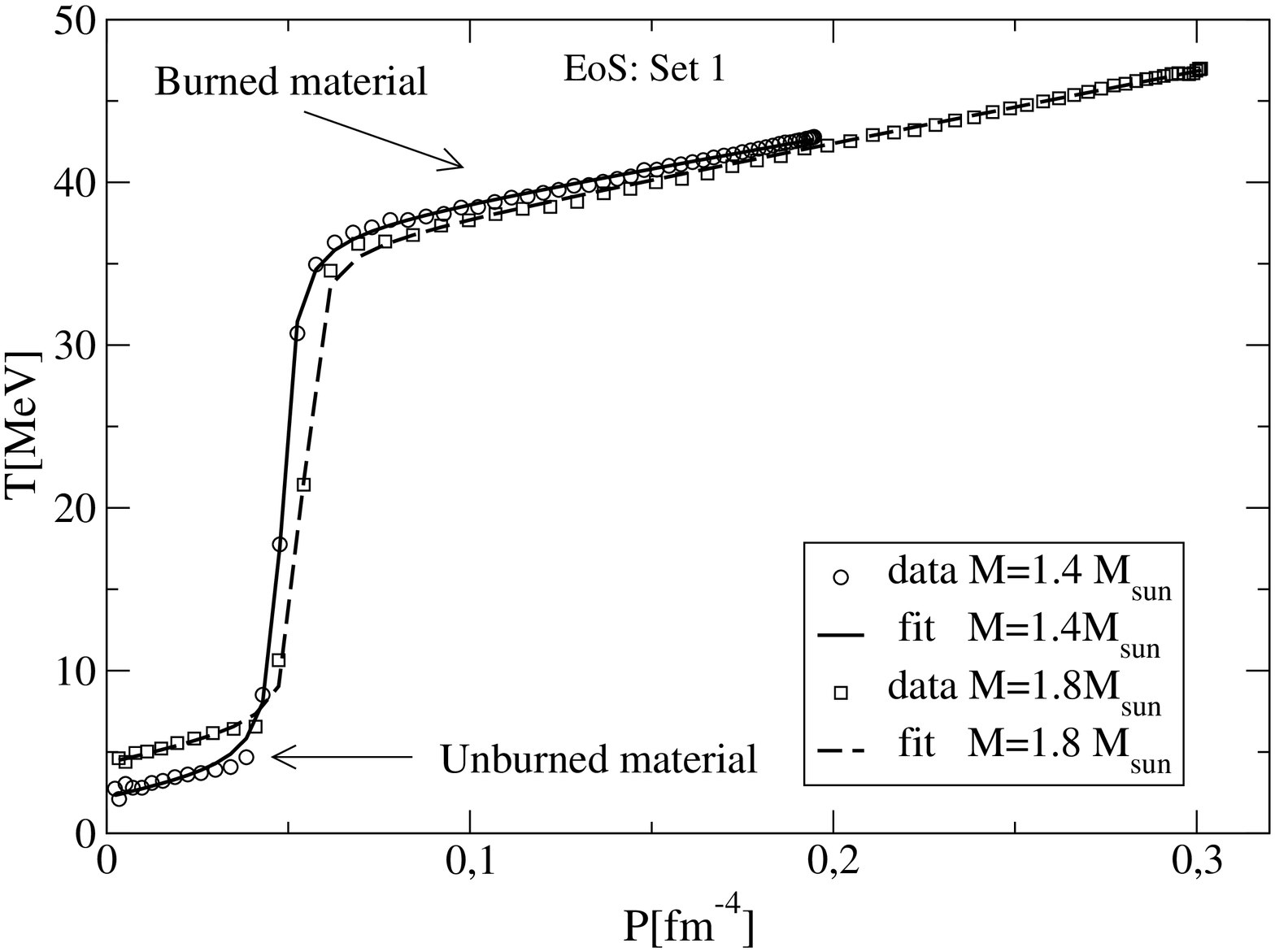,height=8.5cm,width=6cm,angle=-90}
\caption{Temperature as a function of the pressure after combustion. The two lines correspond to the 
numerical output of the simulations and the parametrization which are used for the diffusion calculations.}
\end{centering}
\end{figure}

\begin{figure}
\vskip 0.5cm
\begin{centering}
\epsfig{file=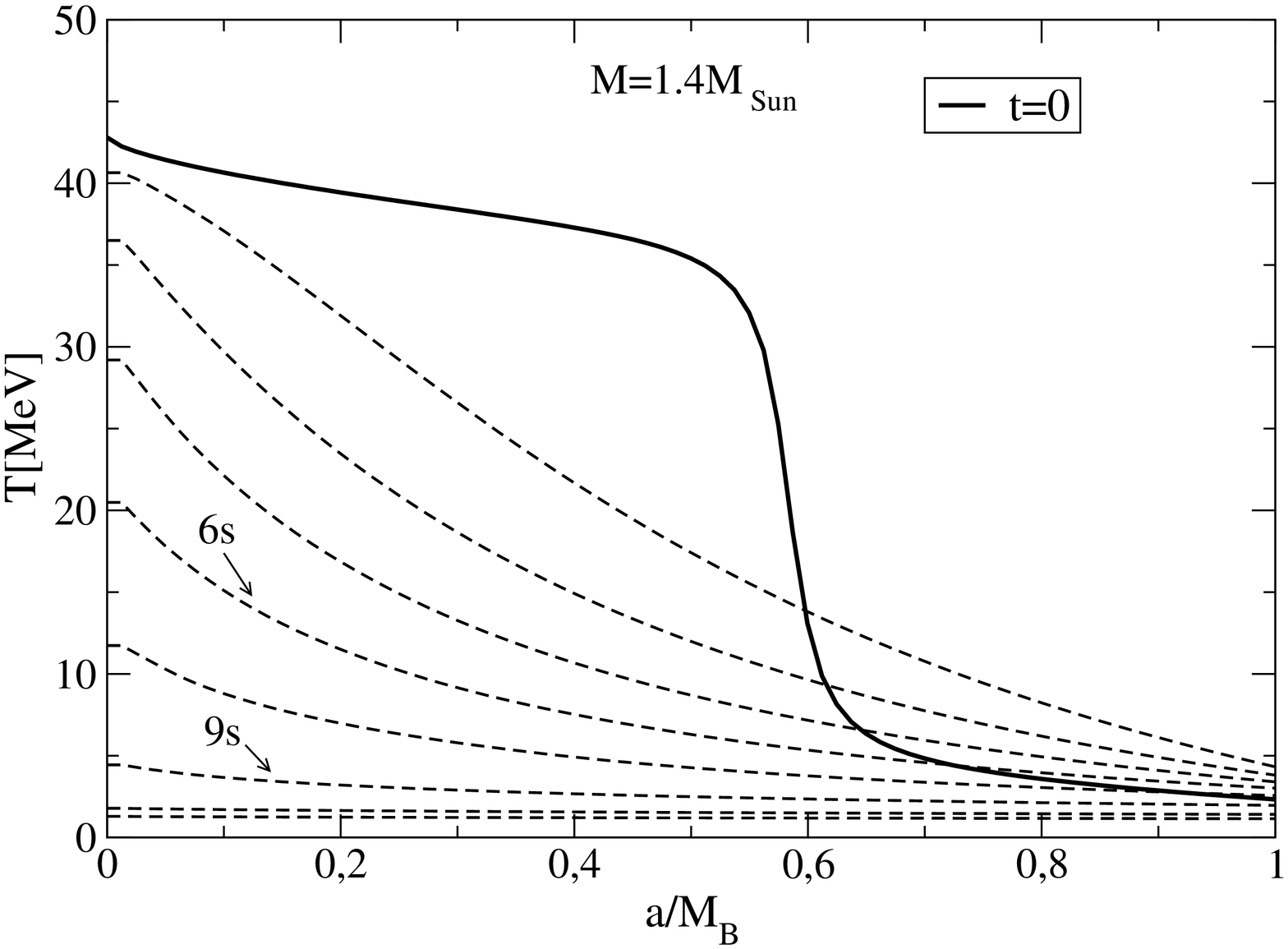,height=8.5cm,width=6cm,angle=-90}
\caption{Temperature evolution as a function of enclosed baryonic mass for a $1.4 M_{\odot}$ star.}
\end{centering}
\end{figure}

\section{Cooling of the strange quark star}
The combustion process lasts typically a few milliseconds. On the other hand, the
diffusion of neutrinos occurs on typical time scales of the order of
tens of seconds. It is therefore a reasonable approximation to
consider the two processes separately: One can use the output of the
combustion simulation as an initial condition for the diffusion
simulation. The diffusion approximation, usually adopted for the study
of the evolution of protoneutron stars, is based on the assumption
that neutrinos are locally in chemical and thermal equilibrium with
baryonic matter and that their motion is driven by the gradients of
chemical composition and temperature within the star. This
approximation is valid if the neutrino's mean free path is much smaller
than the size of the star, which turns out to be the case in
protoneutron star matter and also in the system we are studying here. Of
course, a better but much more complex approach would be to use a Boltzmann transport code
\cite{Huedepohl:2009wh,Fischer:2009af}, which must be necessarily used
in supernova simulations in which neutrinos propagate through very low
density material in the outer layers of the progenitor star.

There is an important difference between protoneutron star matter and
the matter after the combustion of a neutron star into a quark star.
In the first case, apart from the high temperature, matter is also lepton
rich ($Y_L \sim 0.4$), and one can define a flux of net lepton number
which diffuses inside the star and drives the deleptonization. After
this process the stellar matter will be in beta equilibrium.  The energy flux
driven by neutrinos determines instead the cooling of the protoneutron
star (see the simulations of protoneutron star evolution within the
diffusion approximation in
Refs.\cite{1986ApJ...307..178B,janka1,Prakash:1996xs,Pons:1998mm,Pons:2000xf,Pons:2001ar,Pagliara:2010na}).

In our case, before the combustion, the matter of
the neutron star was already in beta equilibrium, and the formation
of the quark phase leads mainly to a strong reheating of matter. We
can thus assume that the most important process for the evolution of
the newly born quark star is the diffusion of the neutrino energy and
we will use only the diffusion equation associated with the transport
of energy \footnote{In principle, absorption processes of neutrinos by
  down quarks could lead to a ``re-leptonization'' of the star, i.e. a
  temporal window in which, probably within the outer layers of the
  star, matter changes its chemical composition. We retain this
  possibility as a future development of this work.}.

An important point must be emphasized here: Within the hydrodynamical
description of the combustion process, the conservation of the total
lepton number is not imposed. The conversion indeed involves the
beta stable nucleonic phase and the beta stable quark phase but
the electron fraction in the nucleonic phase although small, $Y_e \sim
0.1$ at the center of the star, is much larger than the electron
fraction within the quark phase (typically of the order of $10^{-5}$
by using an estimate for the electron chemical potential to be
$\mu_e=m_s^2/4\mu$ \cite{Alford:2007xm}).  In principle, to impose
lepton number conservation, one would need to add another conservation
equation to the hydrocode and use a table for strange quark matter
with another independent variable, i.e. the electron
fraction. Clearly, such a procedure would lead to a different initial
condition for the diffusion calculation.

Let us discuss now the equations that we solve to simulate the
diffusion process. Apart from the cooling of the star caused by
neutrino diffusion, one has also to consider the readjustment of the
structure of the star due to the decreasing of the temperature.
In the standard procedure, the diffusion equation is coupled with the structure
equations which provide, at each time step, the hydrostatic
configuration of the star.
The energy diffusion equation and the
structure equations read \cite{janka1}
\begin{eqnarray}
\frac{\mathrm{d}}{\mathrm{dt}}\frac{\epsilon_{tot}}{n_b}+P\frac{\mathrm{d}}{\mathrm{dt}}\frac{1}{n_b}
&=&-\frac{\Gamma}{n_b r^2\mathrm{e}^\Phi}\frac{\partial}{\partial r}
\left(\mathrm{e}^{2\Phi}r^2\left(F_{\epsilon,\nu_e}\right.\right.\\ \nonumber
&+&\left.\left. F_{\epsilon,\nu_{\mu}}   \right)\right)\\
\frac{dP}{dr}&=&-(P+\epsilon_{tot})\frac{m+4\pi r^3P}{r^2-2mr}\\
\frac{dm}{dr}&=&4\pi r^2\epsilon_{tot}\\
\frac{da}{dr}&=&\frac{4\pi r^2n_b}{\sqrt{1-2m/r}}\\
\frac{d\Phi}{dr}&=&\frac{m+4\pi r^3P}{r^2-2mr}
\end{eqnarray} 
where $\epsilon_{tot}$, $n_b$, and $P$ are the total energy density,
baryon density, and pressure of matter, respectively, $\Gamma=\sqrt{1-2m(r)/r}$ where
$m(r)$, is the gravitational mass enclosed within a radius $r$, $a$ is
the enclosed baryonic mass and $\mathrm{e}^\Phi=\sqrt{g_{00}}$. The
fluxes associated with the transport of energy by electron
neutrinos and antineutrinos and muon and tau
neutrinos and antineutrinos are:
\begin{eqnarray}
F_{\epsilon,\nu_e}&=&-\frac{\lambda_{\epsilon,\nu_e}}{3}\frac{\partial \epsilon_{\nu_e}}{\partial r}\\
F_{\epsilon,\nu_\mu}&=&-\frac{\lambda_{\epsilon,\nu_{\mu}}}{3}\frac{\partial \epsilon_{\nu_{\mu}}}{\partial r}
\end{eqnarray}
where $\epsilon_{\nu_e}$ and $\epsilon_{\nu_{\mu}}$ are the energy
densities of electron and muon neutrinos, respectively , while $\lambda_{\epsilon,\nu_e}$
and $\lambda_{\epsilon,\nu_\mu}$ represent spectral averages of the
mean free paths of $\nu_e$ and $\nu_\mu$, respectively (the transport of $\nu_\mu$
and $\nu_\tau$ is treated as usual on the same footing).

To solve the equations previously introduced one needs, besides the
initial condition on the temperature taken from the hydrodynamical simulations,
the boundary conditions for the temperature at the center of the star
and at the star's radius $R$. At $r=0$, one imposes that the energy fluxes
$F_{\epsilon,i}$ vanish due to symmetry reasons. At the surface we
assume, as in \cite{janka1}, the neutrino radiation to stream off into
vacuum and impose the following conditions on the values of the
fluxes: $F_{\epsilon,i}=\beta \epsilon_i$ where $\beta$ is a geometric
factor which measures the degree of anisotropy of the radiation field
at the surface. Values $\sim 0.5$ are usually considered for this
parameter \cite{janka1,Pons:1998mm}.

Another important ingredient is of course the EOS.  Here a second main
assumption for this calculation has to be done: The simulations of
\cite{Herzog:2011sn} have shown that, although in principle the whole
star should convert into a strange quark star because of the
hypothesis of absolutely stable quark matter, the combustion actually
stops before the full conversion is reached. Most probably the
conversion could go on via diffusion of strange quarks into the
hadronic phase, as computed in \cite{1987PhLB..192...71O}, and of
neutrons into the quark phase, a process
that probably is exothermic as pointed out in \cite{Lugones:1994xg}. We here assume
that the conversion was indeed complete and we use therefore only the
EOS of quark matter for solving the diffusion equation. While
introducing a layer of unburned nucleonic matter would not alter
substantially the initial luminosity of the neutrino signal (at low
densities the mean free paths of quark matter and nucleonic matter are
both large and neutrinos are almost untrapped), the further conversion
of the layer of the unburned material represents a source of energy
which will necessarily prolong the neutrino signal. A model for the
burning of the nucleonic material left after the hydrodynamical
combustion would be needed.

Finally, we must specify the neutrino mean free paths.  As explained
before, for the system we are investigating here, cooling is likely to
be the most important process, while the deleptonization (or the
releptonization) should be subdominant. Within this assumption the
most important reaction between neutrinos and quark matter is the
scattering of nondegenerate neutrinos whose inverse mean free path
reads \cite{Steiner:2001rp}
\begin{equation}
\frac{\sigma_S}{V}=\frac{G_F^2E_\nu^3\mu_i^2}{5\pi^3}
\end{equation}
where $G_F=1.17$ GeV$^{-2}$ is the Fermi constant, $E_\nu$ is the
energy of the neutrino or antineutrino and $\mu_i$ is the chemical
potential of the particle involved in the scattering (up, down,
or strange quarks).
Here, instead of calculating the spectral averages of the mean free paths, we evaluate
the mean free path at $E_\nu=\pi T$, i.e., at the mean energy of neutrinos in thermal equilibrium.

With this setup we can now numerically solve Eqs. (1)-(5) (see
\cite{Pagliara:2010na} for the discussion of the algorithm used).
Results for the temperature profiles as functions of the fraction of
enclosed baryonic mass (the enclosed baryonic mass $a$ over the total
baryonic mass $M_B$) are shown in Figs.3 and 4 for the $1.4$ and $1.8
M_{\odot}$ cases, respectively.  The initial temperature profile
presents a steep gradient as resulting from the combustion
simulations. This steep gradient leads, within the diffusion
approximation we are adopting here, to a strong flux of heat which
rapidly smoothes the temperature profile and after a few seconds the
central temperature drops to half of its initial value. After $\sim
10$s and $\sim 15$s in the two cases, respectively, the star has
reached a uniform temperature of about $1$ MeV and neutrinos can freely
stream within the star, the standard long term cooling sets in at this point. Notice that the
higher the mass of the star, the longer the time needed to cool the
star. The most important result of this paper is shown in Fig.~5, where we
display the total neutrino and antineutrino (summed over the flavors)
luminosity as a function of time.  The luminosity is calculated as
\cite{Pons:1998mm} 

\begin{equation}
L=-e^{2\phi}4\pi r^2 \frac{\lambda_{\epsilon_\nu}}{3}\frac{\partial \epsilon_\nu}{\partial
  r}\mid_{r=R}
\end{equation} 

Starting with an initial value of $\sim 3\times
10^{52}$ erg/s, it drops after $10-15$s to $10^{50}$ erg. The
initial value of the luminosity is very similar to the typical values
of protoneutron stars thus making the process studied here as 
interesting as protoneutron stars from the phenomenological point of
view. The obtained signal lasts, however, shorter by a factor $3-4$ than the
protoneutron star signal (for which the luminosity of $10^{50}$ erg
is reached after about $60$s).  The shorter evolution is attributed to
different aspects: i) The neutrino mean free paths in quark matter are
larger than in nucleonic matter \cite{Steiner:2001rp}, and 
ii) in protoneutron stars the deleptonization 
process mentioned before significantly contributes to the emission.
This study represents a
first estimate of the neutrino signal emitted after the conversion of
a neutron star into a strange star. Clearly, more detailed studies
with a better treatment of the neutrino transport, an EOS extended to
nonbeta stable matter to include the lepton number conservation
during the burning and its transport within the star after the
formation of quark matter will improve the estimates. As explained before, the burning of the
material left after the hydrodynamical combustion could represent an
important source of energy which would then prolong the neutrino
emission. 
\begin{figure}
\vskip 0.5cm
\begin{centering}
\epsfig{file=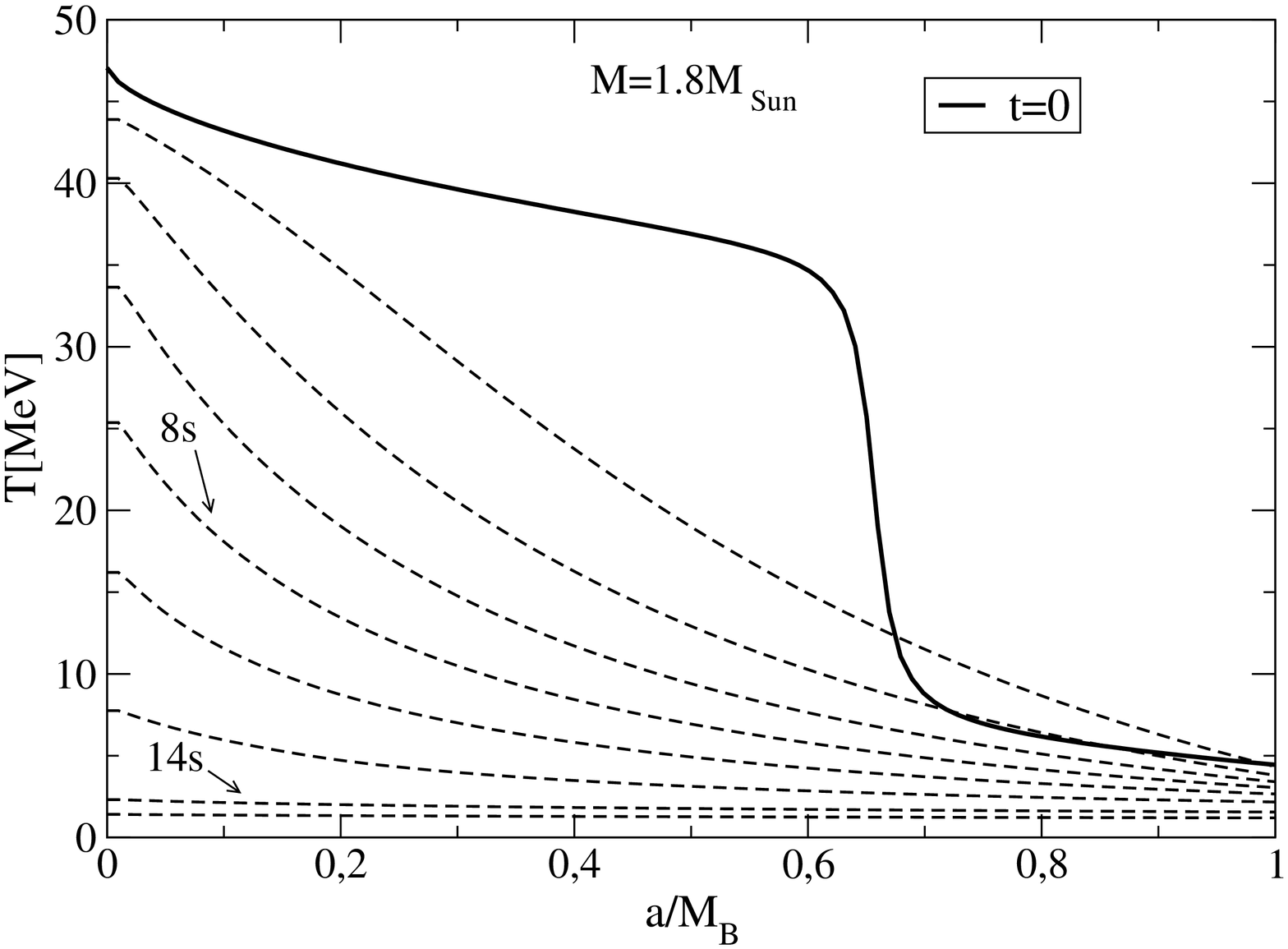,height=8.5cm,width=6cm,angle=-90}
\caption{Temperature evolution as a function of enclosed baryonic mass for a $1.8 M_{\odot}$ star.}
\end{centering}
\end{figure}
\bigskip
In \cite{Niebergal:2010ds} one-dimensional hydrodynamic
simulations of the combustion flame, including neutrino
emission and strange quark diffusion have been conducted:
It turns out that neutrino cooling can halt the burning
interface. It would be interesting to extend these calculations
by use of a three-dimensional hydrodynamic code.
Finally, we note that there are substantial uncertainties in the models 
usually adopted for describing the high density nuclear matter
equation of state. These affect
the results displayed in Fig.~5, which are to be taken as an order of magnitude calculation. It is, however, interesting 
to note that the estimate of the total energy released in the conversion, $\sim 10^{53}$ erg, is consistent
with other estimates obtained by using equations of state for nuclear matter including hyperons;
see, for instance, \cite{Bombaci:2004mt}.

\begin{figure}[ptb]
\vskip 0.5cm
\begin{centering}
\epsfig{file=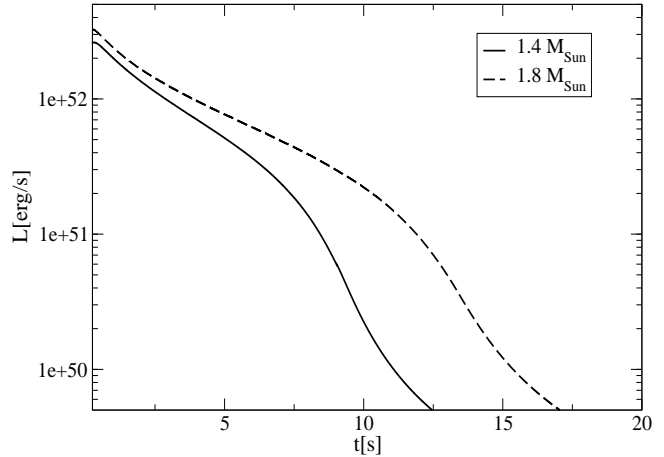,height=8.5cm,width=6cm,angle=-90}
\caption{Total neutrino and antineutrino luminosity as a function of time.}
\end{centering}
\end{figure}
\bigskip

\section{Discussion and conclusions}
From the previous results on the neutrino luminosities one can easily
estimate the total energy released by the conversion process to be of
the order of $10^{53}$ erg. It is thus powerful enough
to be compared with the
energy released within the most violent and, to some extent, still
mysterious explosions of the Universe, i.e., SNe and long GRBs. It is then
clearly tempting to associate at least {\it some of these explosions}
to the formation of a strange star (the first proposals about this connection were presented many years ago 
\cite{DeRujula:1987pg,Cheng:1995am}, see also the more recent Refs. \cite{Bombaci:2000cv,Ouyed:2001cg,Berezhiani:2002ks} ). In particular, the formation of
strange quark matter (regardless of its absolute stability) could
provide an additional energy injection which triggers the explosion of
core collapse supernovae
\cite{Gentile:1993ma,Drago:1997tn,Sagert:2008ka,Fischer:2010wp,Drago:2008tb,Pagliara:2009dg}.
None of the presently available simulations could produce
explosions for high mass progenitors (with masses larger than $\sim 20
M_{\odot}$), and the possible appearance of quark matter could help
in solving this problem.

Apart from the energy released in the birth of a strange star, also the
temporal delay with respect to the collapse of the progenitor star and the birth of the neutron star
could possibly explain some of the puzzling observations connected to
SNe and GRBs. In Ref.~\cite{DeRujula:1987pg} a two-neutrino-burst
scenario is proposed for the neutrino signal of SN1987A: the data of
the LSD detector would indicate a burst of neutrinos that occurred $\sim 5$
h before the well-known K2, IMB and Baksan neutrino events. The second
burst, suggests the author, could be, for instance, associated with the birth of a
strange star.  

It is widely accepted that long GRBs are phenomena connected with the
collapse of massive stars and that they are intimately connected to SNe. In
some cases, a sizable temporal delay (ranging from hours to years)
between a SN and the subsequent GRB was inferred from the data (see
\cite{Berezhiani:2002ks}). In those cases the second explosive event,
i.e. the GRB, could be associated with the conversion of a neutron
star into a star containing quark matter as proposed in
\cite{Berezhiani:2002ks}.

The puzzling observations mentioned before (the LSD neutrino signal
and the long time delay between SN and GRB) have been, however, under debate
for many years and none of them is considered to be statistically
robust. Therefore they do not provide a clear proof
of the existence of quark matter in astrophysical systems.  A more
direct and clean analysis can instead be performed just by considering
the light curves of the prompt emissions of GRBs: It seems again that,
at least in some cases, after a first burst a second burst occurs
which is delayed by up to hundreds of seconds with respect to the
first \cite{Drago:2005rc}. Between the two bursts a quiescent time is present during which
it is likely that the inner engine is not active. In
\cite{Drago:2005rc}, by performing a statistical analysis using the
sample of GRB light curves of the BATSE satellite, 
hints are presented in favor of the interpretation of long quiescent times as periods
during which the inner engine is indeed dormant. A spectacular event
of this type was recently detected by the Swift satellite
\cite{Zhang:2011vk}: The second burst is $11$ min delayed with respect
to the first one. Such a long quiescent time challenges 
popular models for the GRB inner engine, i.e., the collapsar model
\cite{Woosley:1993wj} and the protomagnetar model
\cite{Metzger:2010pp}. In \cite{Zhang:2011vk}, the following
scenario is proposed: The first burst is generated by a rapidly
rotating magnetar, and the second burst is due to a delayed collapse of
the star into a black hole (if enough mass accretes onto the magnetar,
about $1 M_{\odot}$). Recent numerical simulations of the accretion
induced collapse of a neutron star indicate, however, 
that these events could be sources of short GRBs instead of the long ones \cite{Giacomazzo:2012bw}.
Here we speculate that those double bursts could be instead related to
the conversion of a neutron star to a strange star.  Within the
protomagnetar model of long GRBs, the source of energy is provided by
the rotational energy of the star and for the prompt emission, in addition to
the spin down rate, also the neutrino wind released by the hot surface
of the star is crucial: A high neutrino luminosity implies a large
value for the mass loss rate which inhibits the mechanism at the
origin of the gamma radiation. Only when the neutrino luminosity drops
to a critical value, in the untrapping regime, is the prompt emission of
the GRB realized. 
The strange star at birth could then, in principle, generate a new
burst. The quiescent time would correspond, in this scenario, to the
time needed to trigger the conversion process (for instance, because
of the spinning down, the central density increases and at some point
nucleation can start as computed in \cite{Yasutake:2004kx}). A detailed numerical study of this possibility
is an important outlook of this work.

Clearly, the main motivation of this paper is to show that the
conversion process of a neutron star into a strange star generates a
strong neutrino signal which is relevant from the phenomenological point
of view.  Being the first quantitative study in this problem, several
assumptions had to be adopted. For studying quantitatively the phenomenological consequences of 
this process it is important to
improve our calculation especially by introducing the chemical
potential of electron neutrinos and the lepton number diffusion
equation. Modeling the burning of the material left after the
combustion would also be important for obtaining a better estimate of the
duration of the neutrino signal.

Finally, testing our theoretical results by means of a direct
neutrino detection is clearly very difficult: If such processes
really occur in the Universe, their rate is probably significantly
lower than that of core collapse SN events, 
therefore making a detection highly improbable. On the other hand, we have plenty of
data on long GRBs: We have suggested that these data already contain
some interesting information which would possibly indicate that strange
quark matter is really formed in compact stellar objects.

We thank A. Drago for valuable discussions.  G.P. acknowledges
financial support from the Italian Ministry of Research through the
program \textquotedblleft Rita Levi Montalcini\textquotedblright.  The
work of F.K.R. is supported by the Deutsche Forschungsgemeinschaft
(DFG) via the Emmy Noether Program (RO 3676/1-1), and by the ARCHES
prize of the German Ministry of Education and Research (BMBF).

\bigskip


\end{document}